\documentclass[aps,prb,superscriptaddress,twocolumn,preprintnumbers,amsmath,amssymb]{revtex4}

\usepackage{graphicx}
\usepackage{epstopdf}
\usepackage{longtable}
\usepackage{bm}
\usepackage{multirow}
\usepackage{hyperref}
\usepackage{array}
\usepackage{booktabs}
\usepackage{amssymb}
\usepackage{amsmath}

\begin{document}

\title{Coupling of phonons with orbital dynamics and magnetism in CuSb$_2$O$_6$}

\author{D. T. Maimone}
\affiliation{``Gleb Wataghin'' Institute of Physics, University of Campinas - UNICAMP, Campinas, S\~ao Paulo 13083-859, Brazil}

\author{A. B. Christian}
\affiliation{Department of Physics, Montana State University, Bozeman, Montana 59717, USA}

\author{J. J. Neumeier}
\affiliation{Department of Physics, Montana State University, Bozeman, Montana 59717, USA}

\author{E. Granado}
\affiliation{``Gleb Wataghin'' Institute of Physics, University of Campinas - UNICAMP, Campinas, S\~ao Paulo 13083-859, Brazil}

\begin{abstract}

Strongly interacting phonons and orbital excitations are observed in the same energy range for CuSb$_2$O$_6$, unlocking a so-far unexplored type of electron-phonon interaction. An orbital wave at $\sim 550$ cm$^{-1}$ softens on warming and strongly interferes with a phonon at $\sim 500$ cm$^{-1}$, giving rise to a merged excitation of mixed character. An electronic continuum grows on warming to the orbital ordering temperature $T_{OO}$=400 K, generating an important phonon decay channel. This direct and simultaneous observation of orbital and vibrational excitations reveals details of their combined dynamics. In addition, phonon frequency anomalies due to magnetic correlations are observed below $\sim 150$ K, much above the three-dimensional magnetic ordering temperature $T_N^{3D}=8.5$ K, confirming one-dimensional magnetic correlations along Cu-O-O-Cu linear chains in the paramagnetic state. 

\end{abstract}


\maketitle

\section{Introduction}
Remarkable collective effects in condensed matter physics such as superconductivity and colossal magnetoresistance originate from the coupling between electronic degrees of freedom and atomic displacements. In these cases, not only the electronic and lattice ground states may be intertwined, but also the corresponding excitations may in principle be coupled and have mixed character. Among the various possible mechanisms for electron-phonon interaction, the orbital-lattice coupling through the Jahn-Teller (JT) effect of transition-metal ions with half-filled $e_g$ level is a case of paramount importance. Since the electronic splitting of $e_g$ levels is normally of the order of 1-2 eV, much above the phonon range ($\lesssim 0.1$ eV), the first-order orbital and phononic excitations themselves usually do not mix, maintaining their pure character, while vibronic second-order excitations may be accessed through the well known Franck-Condon mechanism. A distinct and highly interesting situation would arise if the energies of the $e_g$ orbital excitations could be tailored to match the phonon energy scale, thus favoring strongly interacting orbital and phononic first-order excitations.

A promising material to exhibit this so-far unexplored physics is CuSb$_2$O$_6$, with a small Jahn-Teller distortion of CuO$_6$ octahedra \onlinecite{Giere,Nakua} that favors a competition between the Cu $3d_{3z^2-r^2}$ and $3d_{x^2-y^2}$ $e_g$ levels \onlinecite{Kasinathan}. In this work, we present a Raman scattering study of CuSb$_2$O$_6$ and report strong evidence of low-energy orbital dynamics that interact strongly with phonons, opening a new avenue to study the interplay between electronic and vibrational degrees of freedom in systems with significant orbital-lattice interaction. Also, anomalous phonon shifts below $\sim 150$ K reveal magnetic correlations along the Cu-O-O-Cu super-superexchange path in the {\it ab}-plane of the trirutile structure \onlinecite{Koo,Whangbo,Kasinathan}, confirming the detailed origin of the one-dimensional (1D) magnetism observed in this material \onlinecite{Koo,Whangbo,Kasinathan,Nakua,Nakua2,Yamagushi,Kato,Kato2,Prokofiev,
Heinrich,Torgashev,Gibson,Rebello,Herak,Wheeler,Prasai}. This analysis proves that spin-phonon spectroscopy can provide decisive information on the dominant exchange paths in magnetic materials, complementing other magnetic spectroscopy techniques.

\section{Experimental Details}
The CuSb$_2$O$_6$ crystal was grown by chemical vapor transport \onlinecite{Prokofiev,Rebello}. A natural grown {\it ab} face was identified by Laue x-ray diffraction and mounted on the cold finger of a closed-cycle He cryostat with a base temperature of 20 K for the Raman experiment, which was performed in quasi-backscattering geometry using the 488.0-nm line of an Ar$^{+}$ laser with a spot focus of $\sim 100$ $\mu$m. A triple 1800 mm$^{-1}$ grating spectrometer equipped with a $L$N$_2$-cooled multichannel CCD detector was employed. The damping parameters $\Gamma(T)$ were obtained after deconvolution of an instrumental contribution $\Gamma_{inst} = 3.4$ cm$^{-1}$. Partial depolarization effects were observed in our Raman spectra, which is ascribed to the imperfect quasi-backscattering geometry of our experiment and a possible birefringence effect as noted in Ref. \onlinecite{Maimone}. This effect did not allow for a reliable symmetry analysis of the electronic Raman signal reported here. 

\section{Results, Analysis and Discussion}

\begin{figure}
	\includegraphics[width=0.4 \textwidth]{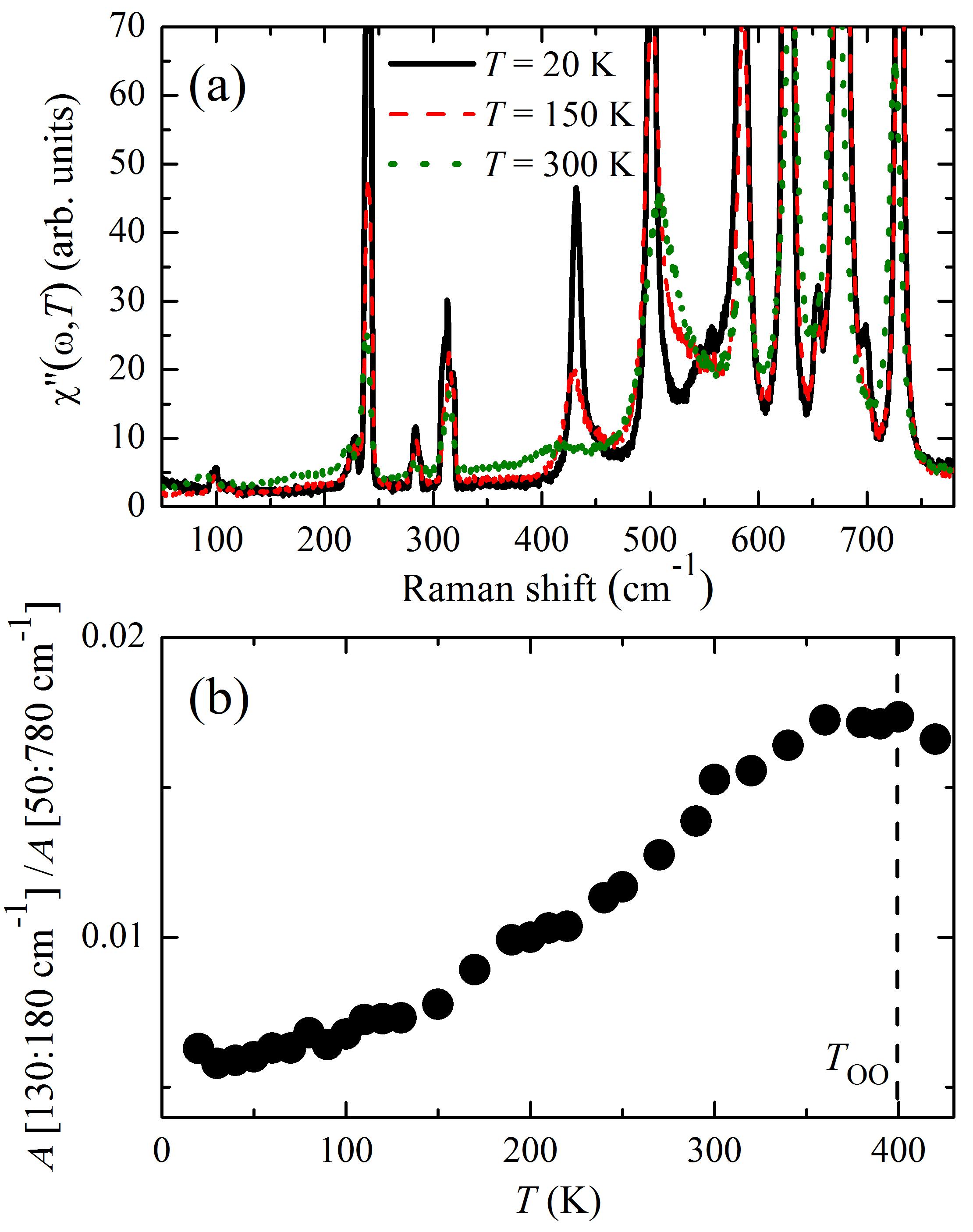}
	\caption{\label{spectraraw} (a) Raman response $\chi''(\omega,T)$ at selected temperatures. (b) Temperature dependence of the area $A$ of $\chi''(\omega,T)$ computed between 130 and 180 cm$^{-1}$ normalized by $A$ computed between 50 and 780 cm$^{-1}$. The vertical dashed line indicates the orbital ordering temperature $T_{OO}=400$ K.} 
\end{figure}

\begin{figure}
	\includegraphics[width=0.4 \textwidth]{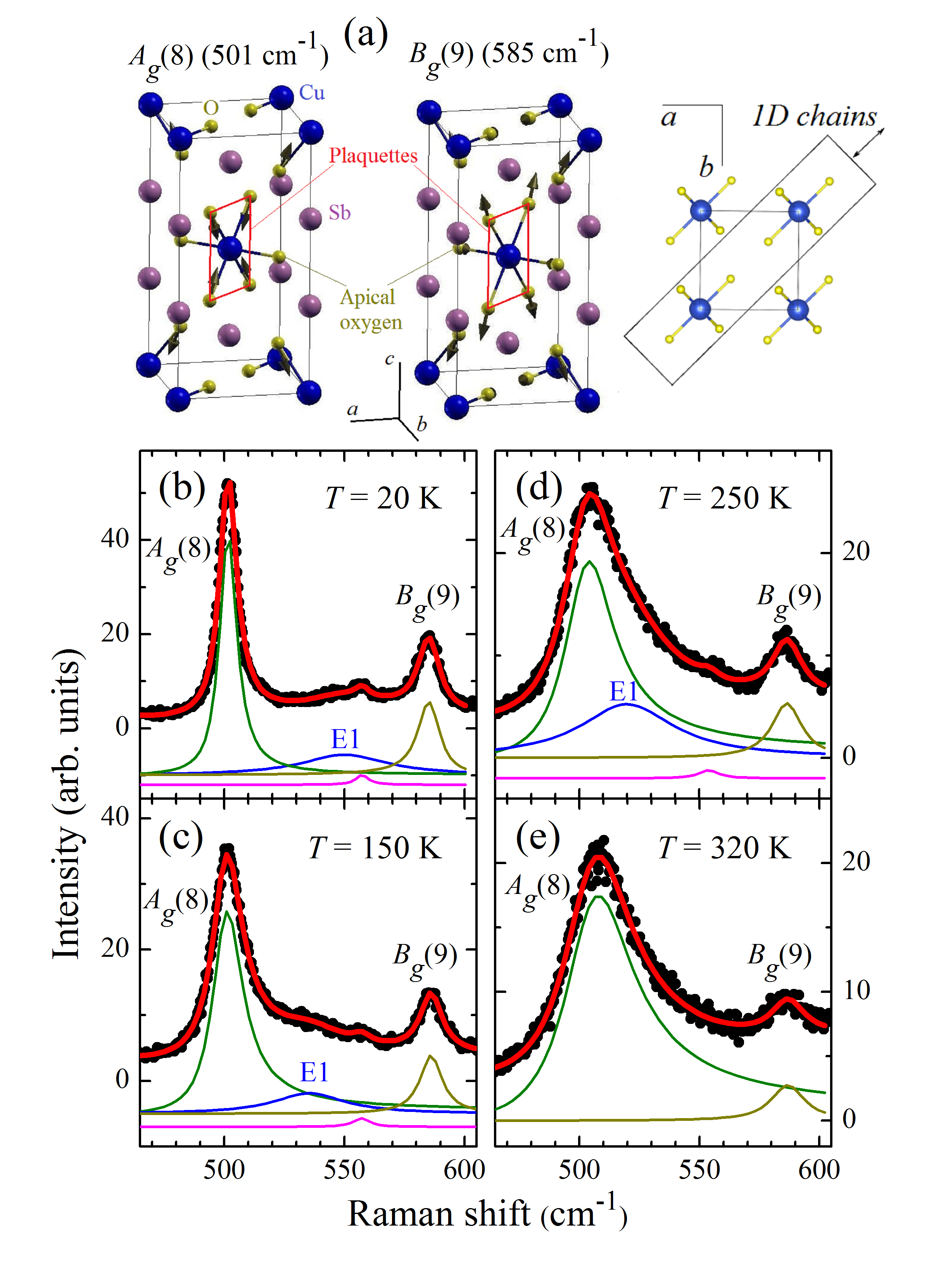}
	\includegraphics[width=0.4 \textwidth]{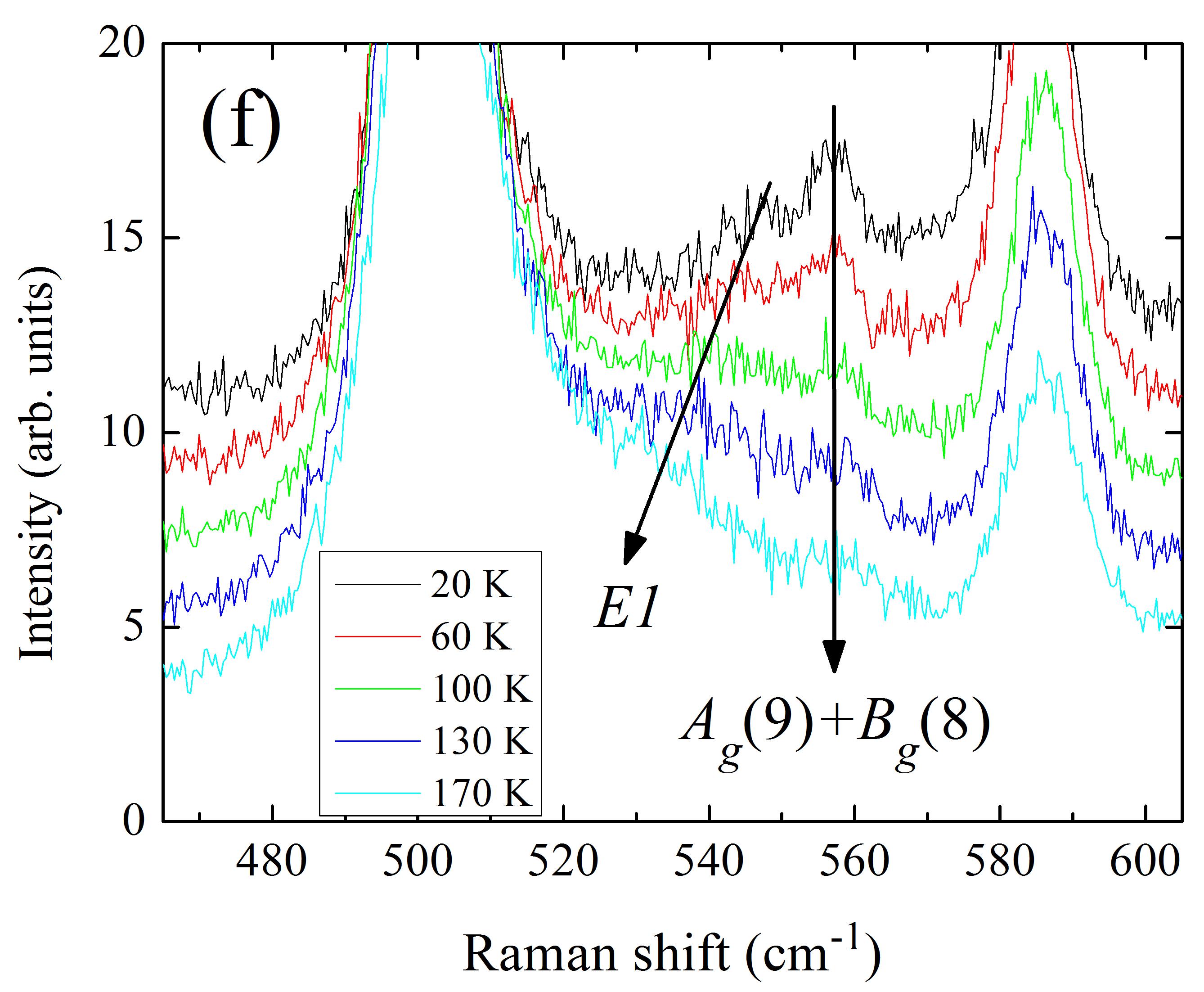}
	\caption{\label{spectrafit} (a) Crystal structure of CuSb$_2$O$_6$ and mechanical representations of the vibrational modes $A_g(8)$ at 501 cm$^{-1}$ and $B_g(9)$ at 585 cm$^{-1}$ \onlinecite{Maimone}. The CuO$_4$ plaquettes and apical oxygen ions as defined in Ref. \onlinecite{Kasinathan}  are indicated. An {\it ab}-plane projection of the CuO$_6$ octahedra with Cu and apical oxygen ions at $z=0$ is also shown, identifying the super-superexchange path leading to 1D magnetic chains \onlinecite{Kasinathan,Koo,Whangbo}. (b-e) Selected portion of the Raman spectra at selected temperatures. Closed symbols: experimental data; red line: fit to a four-peak (b-d) or two-peak (e) function (see text). The peak decomposition is indicated in solid lines at the bottom of each panel, and the electronic $E1$ peak and phonon $A_g(8)$ and $B_g(9)$ peaks are explicitly noted. (f) Zoom out of the same spectral region at selected temperatures, highlighting the $E1$ peak and the weak $A_g(9)+B_g(8)$ phonon peak. The spectra in (f) were vertically translated for clarity.}
\end{figure}

\begin{figure}
	\includegraphics[width=0.45 \textwidth]{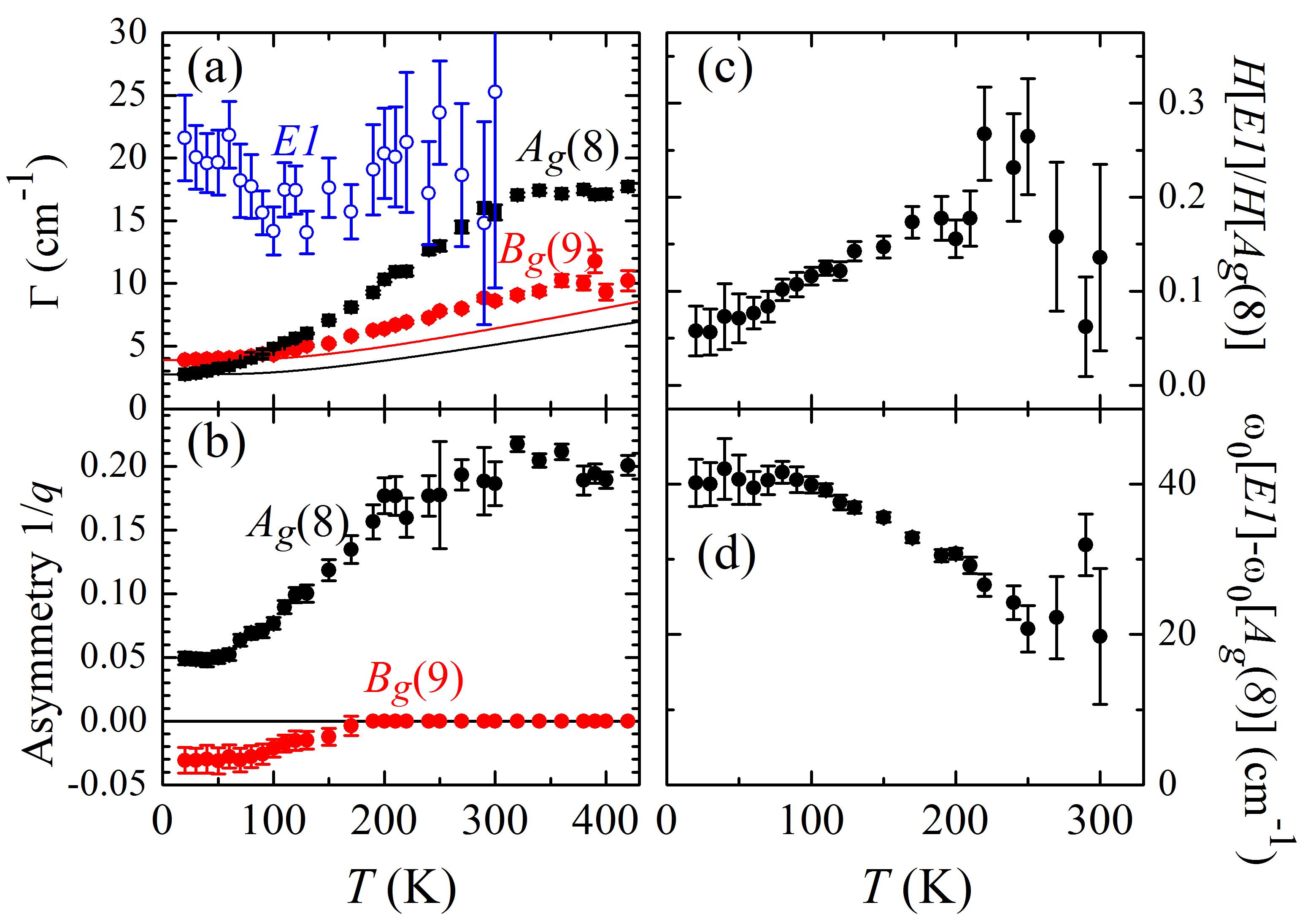}
	\caption{\label{parameters1} Temperature-dependence of (a) damping constant $\Gamma$ of the $A_g(8)$, $B_g(9)$, and $E1$ peaks displayed in Figs. \ref{spectrafit}(a)-\ref{spectrafit}(f); (b) asymmetry parameter $1/q$ of the $A_g(8)$ and $B_g(9)$ peaks; (c) height ratio; and (d) peak frequency difference between the $A_g(8)$ and $E1$ peaks.  The black and red lines in (a) are the two-phonon decay $\Gamma(T)$ curves for the $A_g(8)$ and $B_g(9)$ modes, respectively \onlinecite{Klemens,Balkanski}.}
\end{figure}

\begin{figure}
	\includegraphics[width=0.45 \textwidth]{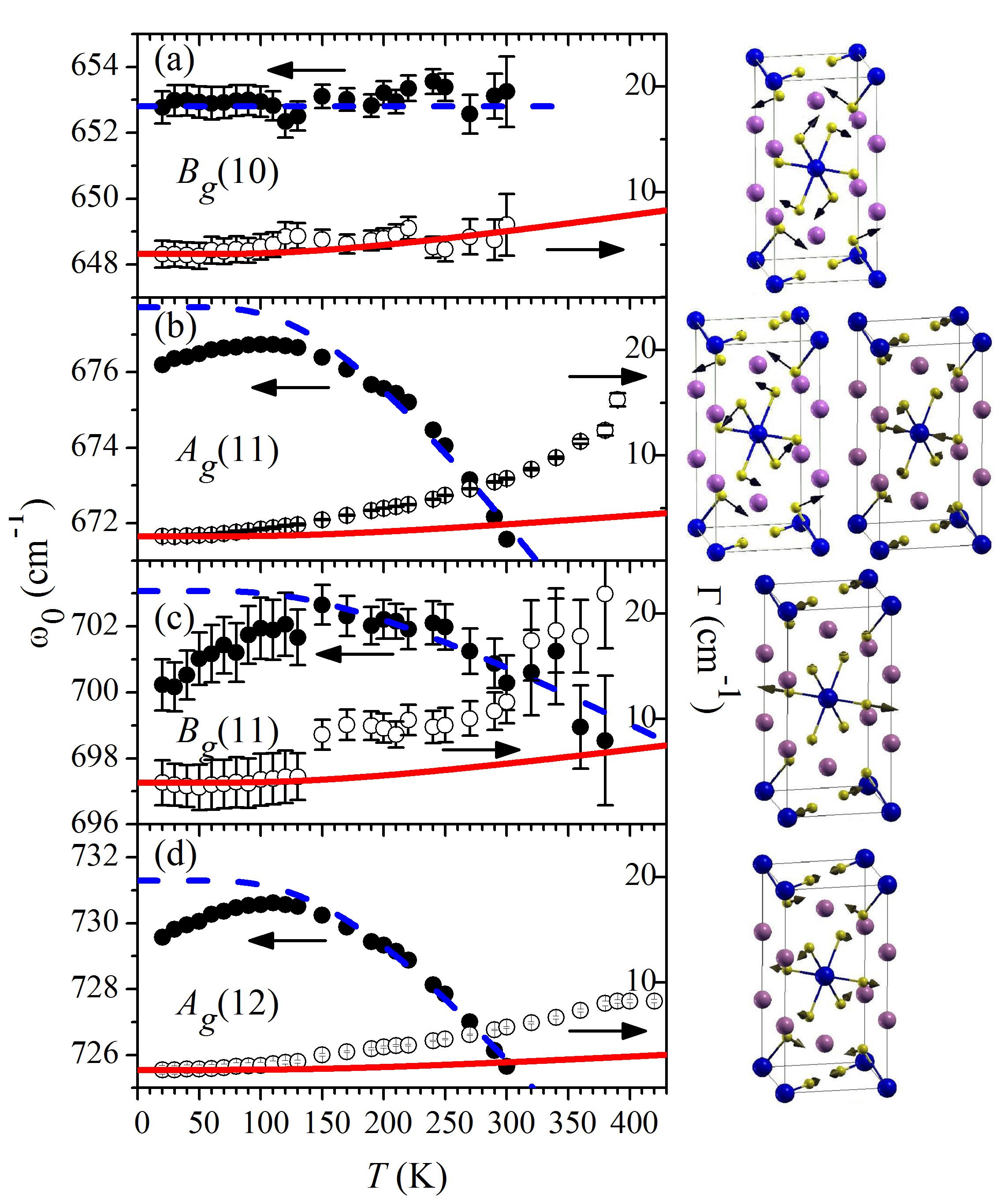}
	\caption{\label{parameters2} Temperature dependence of line positions $\omega_0$ (closed symbols) and damping constants $\Gamma$ (open symbols) of the high-frequency phonon modes $B_g(10)$ (a), $A_g(11)$ (b), $B_g(11)$ (c), and $A_g(12)$ (d). The solid red and dashed blue lines are the two-phonon decay curves for $\Gamma(T)$ and $\omega_0(T)$, respectively \onlinecite{Klemens,Balkanski}. The corresponding mechanical representations are displayed at the right of each panel \onlinecite{Maimone}. The $A_g(11)$ mode is the result of a phonon mixing effect of the two symmetric vibrations displayed in (b) \onlinecite{Maimone}.}
\end{figure}

Figure \ref{spectraraw}(a) shows the unpolarized Raman response $\chi''(\omega)$ at selected temperatures, obtained by correcting the raw intensities by the Bose-Einstein factor. A total of 18 phonon modes are observed at $T=20$ K, close to the expected 24 Raman-active modes in the monoclinic phase, space group $P2_1/n$ ($\Gamma_{Raman} = 12A_g+12B_g$). The complete phonon assignment and a detailed investigation of the soft modes at 98 and 432 cm$^{-1}$ ($T=20$ K) are given in Ref. \onlinecite{Maimone}. A continuum signal superposed to the phonon peaks is also observed up to at least $\sim 700$ cm$^{-1}$, which enhances on warming. Figure \ref{spectraraw}(b) shows the temperature dependence of the raw $\int \chi''(\omega) d\omega$ integrated over a small frequency interval between 130 and 180 cm$^{-1}$, where no phonon peak is observed, normalized by $\int \chi''(\omega) d\omega $ integrated over the whole investigated spectral range ($50 < \omega < 780$ cm$^{-1}$). {\it The Raman continuum enhances on warming up to $T_{OO}$, signaling a clear connection with the orbital degree of freedom.} Such a continuum is highly unusual for insulators. A notable exception is RTiO$_3$ ($R$=rare earth), where a broad continuum was also observed and attributed to liquidlike correlations between spins and orbitals \onlinecite{Reedyk,Ulrich3}.

As shown in Fig. \ref{spectraraw}(a), relevant changes in the Raman signal with temperature are also noted in the spectral window between 500 and 600 cm$^{-1}$. This interval covers the $A_g(8)$ and $B_{g}(9)$ phonon modes represented in Fig. \ref{spectrafit}(a). Figures \ref{spectrafit}(b)-\ref{spectrafit}(e) show the Raman intensities at this spectral range (symbols), where the polarizer was oriented to minimize the $B_g$ signal. Fits of the $A_g(8)$ and $B_{g}(9)$ peaks to asymmetric Breit-Wigner function line shapes $I = H (1+\frac{\omega - \omega_0}{q \Gamma})^2/[1+(\frac{\omega - \omega_0}{\Gamma})^2]$ were performed \onlinecite{Fano}, where $\omega_0$ is the line position, and $H$, $\Gamma$, and $1/q$ are the peak height, damping, and asymmetry parameters, respectively. Also, the inclusion of two additional weak peaks in the fitting model below $T=300$ K significantly improved the quality of the fit. The peak at $\sim 555$ cm$^{-1}$ is associated with quasi-degenerate $A_g(9)+B_g(8)$ modes derived from the $E_g(5)$ phonon of the tetragonal phase \onlinecite{Maimone}. No further Raman-active phonons are expected in this spectral region \onlinecite{Maimone}, and the origin of the additional peak at $\sim 520-550$ cm$^{-1}$ (labeled ``$E1$'') is discussed below. The resulting four-peak fits are shown in Figs. \ref{spectrafit}(b)-\ref{spectrafit}(d) (solid lines), where an additional second-order polynomial background was also included. Figure \ref{spectrafit}(f) shows the evolution of the $E1$ and $A_g(9)+B_g(8)$ peaks with temperature in better detail.

Figure \ref{parameters1}(a) shows the damping constant $\Gamma$ for the $A_g(8)$, $B_{g}(9)$, and $E1$ peaks obtained from our fits. The expected $\Gamma_{2ph}(T)$ curves for the $A_g(8)$ and $B_{g}(9)$ modes according to conventional two-phonon decay, $\Gamma_{2ph}(T) = \Gamma_0[1+2/(e^x-1)]$ with $x \equiv \hbar \omega_0/2 k_B T$ \onlinecite{Klemens,Balkanski}, are also shown. The $B_{g}(9)$ mode shows a stronger damping than the $\Gamma_{2ph}(T)$ curve on warming above $\sim 100$ K. A similar behavior was also found for other modes (see below). Phonon decays into three or more phonons are expected to yield only minor corrections with respect to two-phonon decay in the studied temperature range \onlinecite{Balkanski} and are not able to explain our results. We attribute this effect to {\it phonon decay into the continuum of orbital excitations} evidenced in Fig. \ref{spectraraw} \onlinecite{Khaliullin,Ulrich3}. Intriguingly, the $A_g(8)$ mode shows a $\Gamma(T)$ curve that seems to deviate qualitatively from the behavior of the other phonons. In fact, while this mode is sharper than the $B_{g}(9)$ mode at low temperatures, it becomes much broader above $\sim 100$ K, reaching a nearly $T$-independent $\Gamma \sim 17$ cm$^{-1}$ above $\sim 300$ K. Another remarkable feature of the $A_g(8)$ mode is its asymmetry parameter $1/q$, which shows a fourfold increase from $1/q=0.05$ at $T=20$ K to $1/q=0.20$ above 300 K [see Fig. \ref{parameters1}(b)]. In opposition, a slight negative asymmetry is perceived for the $B_{g}(9)$ mode at $T=20$ K [see also Fig. \ref{spectraraw}(a)], evolving to a fully symmetric line shape above $\sim 170$ K. 

The $E1$ peak shows a large $\Gamma=22(3)$ cm$^{-1}$ at $T=20$ K, which is much broader than any phonon mode at this temperature, remaining with approximately the same $\Gamma$ up to $\sim 300$ K. This peak softens considerably and approaches the $A_g(8)$ mode on warming [see Figs. \ref{spectrafit}(f) and \ref{parameters1}(d)]. Due to this shift and the simultaneous broadening of the $A_g(8)$ phonon, the $E1$ peak becomes superposed with the $A_g(8)$ signal in the raw data above $\sim 200$ K. Still, our fitting model with individual $A_g(8)$ and $E1$ peaks led to well converged results up to $300$ K. The relative height of the $E1$ peak shows a change of behavior  at $\sim 250$ K, becoming weaker above this temperature [see Fig. \ref{parameters1}(c)]. Above $300$ K, this fitting model becomes unstable and a simpler two-peak fitting model was employed [see Fig. \ref{spectrafit}(e)].

{\it The asymmetric line shapes of the $A_g(8)$ and $B_g(9)$ modes demonstrate the presence of an electronic Raman signal that interferes with the phonon signals through a Fano-like mechanism} \onlinecite{Fano,Cerdeira}. The detailed temperature dependence of the asymmetry of these modes indicates that the interfering electronic Raman signal shifts towards the $A_g(8)$ mode and farther away from the $B_g(9)$ mode on warming, thus increasing (decreasing) the magnitude of $1/q$ for the $A_g(8)$ ($B_g(9)$) mode [see Fig. \ref{parameters1}(b)]. Remarkably, this is precisely the behavior of the observed $E1$ peak (see above). {\it This clear correspondence between the phonon asymmetry parameters [Fig. \ref{parameters1}(b)] and position of the $E1$ line [Fig. \ref{parameters1}(d)] allows us to associate the weak $E1$ peak to an electronic excitation}, excluding alternative possibilities such as a disorder-activated first-order phonon or a two-phonon peak, which do not interfere with other normal modes of vibration.

The physical properties of CuSb$_2$O$_6$ severely limit the possible electronic excitations in this energy scale, eliminating any possible ambiguity on the detailed origin of the $E1$ peak. In fact, charge-carrier excitations can be excluded considering the insulating character of this material with the large Mott gap of 2.2 eV \onlinecite{Kasinathan}. Also, this signal is incompatible with magnetic excitations, since even two magnons in a linear chain have a typical energy of $4JS^2 \sim 100$ K ($\sim 70$ cm$^{-1}$) \onlinecite{Nakua,Yamagushi,Kato,Kato2,Prokofiev,Heinrich,
Torgashev,Gibson,Rebello} associated with the energy cost of flipping two neighboring spins, much lower than the energy of $E1$ ($\sim 550$ cm$^{-1}$). On the other hand, {\it the observation of $E1$ is consistent with a collective orbital wave.} In fact, such an excitation can be Raman-active \onlinecite{Ishihara} and was claimed to be observed in titanates \onlinecite{Ulrich,Ulrich2} and vanadates \onlinecite{Miyasaka,Sugai}. Its observation in an $e_g$ orbital system such as a cuprate might seem surprising, since a strong orbital-lattice interaction is believed to suppress the collective orbital waves in favor of single-site crystal-field excitations \onlinecite{Khaliullin}. However, the combination of weak Jahn-Teller distortion of CuO$_6$ octahedra \onlinecite{Nakua,Giere}, a rigid crystal structure, and strongly competing $3d_{3z^2-r^2}$ and $3d_{x^2-y^2}$ orbitals \onlinecite{Kasinathan} favors the formation of low-energy collective orbital excitations. Also, the existence of three distinct Cu-O bond distances in the monoclinic phase below $T_{OO}$ (Ref. \onlinecite{Nakua}) may be associated with a hybrid orbital ground state, corresponding to a linear combination of $3d_{3z^2-r^2}$ and $3d_{x^2-y^2}$ orbitals. A slightly different set of hybrid orbitals and Cu-O distances may yield a metastable state that is quasi-degenerate with the ground state, leading to low-energy orbital excitations, such as proposed for KCuF$_3$ (Ref. \onlinecite{Lee}). In addition to collective orbital excitations, incoherent single-site excitations between the quasi-degenerate orbitals are also allowed. Such excitations are localized in the real space, being therefore broad in {\it k} space, giving rise to broad peaks in the Raman spectrum due to the significant band dispersion of $3d$ $e_g$ levels, at least in the 100 meV scale. Thus, single-site orbital excitations cannot generate the relatively sharp $E1$ Raman peak. In any case, {\it incoherent orbital excitations are the likely source of the electronic continuum revealed in Fig. \ref{spectraraw}}.

The softening of the $E1$ peak upon warming above $\sim 100$ K [see Fig. \ref{parameters1}(d)] is consistent with the scenario drawn above, since the gradual meltdown of the 1D spin-correlated state on warming alters the interorbital superexchange term that contributes to the orbital wave total energy. Above $\sim 300$ K, the $A_g(8)$ and $E1$ peaks are inseparable, consistent with the formation of a single excitation of mixed phononic/electronic character. Notice that the growing damping parameter of the $A_g(8)$ peak on warming seems to be limited above $\sim 300$ K by the intrinsic $\Gamma(E1)$ [see Fig. \ref{parameters1}(a)], favoring our interpretation. {\it Such evolution of two independent phononic and electronic lines merging into a single broad excitation is evidence of a microscopic mechanism for electron-phonon interaction that deserves further investigation.} The decisive feature of the trirutile structure of CuSb$_2$O$_6$ that brought the orbital dynamics to the phonon energy scale is the relative rigidity of the CuO$_6$ octahedra against JT distortions. This is caused by the edge-sharing of these octahedra with undistorted SbO$_6$ octahedra \onlinecite{Giere}, increasing the strain energy penalty of a large Cu$^{2+}$O$_6$ JT distortion. Such rigidity contrasts with the flexibility of perovskite-derived structures to accomodate octahedral distortions. {\it Similar physics as discussed here is likely to be a general occurrence in crystal structures with highly constrained and almost regular $M$O$_6$ octahedra, where $M$ is a JT-active ion.}


In addition to phonon-orbital dynamic effects, signatures associated with the quasi-1D magnetism are clearly evident in the data. Figures \ref{parameters2}(a)-\ref{parameters2}(d) show $\omega_0$ and $\Gamma$ for the four phonon peaks [$B_g(10)$, $A_g(11)$, $B_g(11)$, $A_g(12)$] observed in the spectral region between 650 and 730 cm$^{-1}$. These peaks present symmetric Lorentzian lineshapes at all temperatures. The $A_g(11)$, $B_g(11)$, and $A_g(12)$ modes show stronger damping than the conventional two-phonon decay behavior \onlinecite{Klemens,Balkanski} [see Figs. \ref{parameters2}(b)-\ref{parameters2}(d)], following the same behavior noted above for the $B_g(9)$ mode. In addition, anomalous softenings of 2-3 cm$^{-1}$ are observed for the $A_g(11)$, $B_g(11)$, $A_g(12)$ modes on cooling below $\sim 150$ K, coinciding with the temperature range where the inverse magnetic susceptibility deviates from the Curie-Weiss law, but still much above the three-dimensional (3D) magnetic ordering temperature $T_N^{3D}=8.5$ K \onlinecite{Nakua,Nakua2,Yamagushi,Kato,Kato2,Prokofiev,Heinrich,Torgashev,Gibson,Rebello,Herak,Wheeler}. These phonon shift anomalies are ascribed to the spin-phonon coupling effect in the presence of short-range magnetic correlations \onlinecite{Granado,Flores}. In opposition, the $B_g(10)$ mode does not show such an anomaly. An inspection of the mechanical representations of these modes, displayed in Figs. \ref{parameters2}(a)-\ref{parameters2}(d) \onlinecite{Maimone}, shows that spin-phonon coupling is seen in vibrations with participation of apical oxygen ions, using the nomenclature of Ref. \onlinecite{Kasinathan} [see also Fig. \ref{spectrafit}(a)]. {\it This confirms magnetic correlations along the Cu-O-O-Cu linear chains in the {\it ab} plane, indicating these are the dominant exchange paths as predicted by previous theoretical work} \onlinecite{Kasinathan,Koo,Whangbo}.

\section{Conclusions}

In summary, Raman scattering reveals Cu$^{2+}$ $e_g$ orbital excitations that interact strongly with phonons in CuSb$_2$O$_6$, unlocking a so-far unexplored microscopic mechanism of electron-phonon interaction. The relatively low-energy scale of the orbital excitations in this system is attributed to relatively rigid CuO$_6$ octahedra leading to strongly competing Cu $3d_{3z^2-r^2}$ and $3d_{x^2-y^2}$ orbitals. This physics may be replicated in other crystal structures with highly constrained edge- or face-shared $M$O$_6$ octahedra. In addition, phonon shift anomalies due to spin-phonon coupling confirms 1D magnetic correlations along the Cu-O-O-Cu linear chains in the {\it ab} plane, providing experimental proof of the dominant super-superexchange path theoretically proposed for this system \onlinecite{Kasinathan,Koo,Whangbo}. This analysis proves that spin-phonon spectroscopy can provide direct information on the dominant exchange path in low-dimensional magnetic systems.

\begin{acknowledgments}

This work was conducted with financial support from FAPESP through Grant No. 2012/04870-7, CAPES, and CNPq, Brazil, and the U.S. Department of Energy (DOE), Office of Science, Basic Energy Sciences (BES), under Award No. DE-SC0016156.

\end{acknowledgments}

\end{document}